\documentclass[conference]{IEEEtran}
\usepackage{fancyhdr}
\IEEEoverridecommandlockouts

\usepackage{cite}
\usepackage{amsmath,amssymb,amsfonts}
\usepackage{algorithmic}
\usepackage{textcomp}
\usepackage{array}
\usepackage{graphicx}
\usepackage{booktabs}
\usepackage{multirow}
\usepackage[hyphens]{url} 
\usepackage{xcolor}
\usepackage{balance}
\usepackage{soul}

\usepackage{pdfpages}
\usepackage{pgfplots}
\pgfplotsset{compat=1.17}

\newcolumntype{L}[1]{>{\raggedright\let\newline\\\arraybackslash\hspace{0pt}}m{#1}}
\newcolumntype{C}[1]{>{\centering\let\newline\\\arraybackslash\hspace{0pt}}m{#1}}
\newcolumntype{R}[1]{>{\raggedleft\let\newline\\\arraybackslash\hspace{0pt}}m{#1}}

\def\BibTeX{{\rm B\kern-.05em{\sc i\kern-.025em b}\kern-.08em
    T\kern-.1667em\lower.7ex\hbox{E}\kern-.125emX}}

\newcommand{\aspect}{\ensuremath{\alpha}} 

\newcommand{\rev}[1]{{#1}}

\begin{document}

\includepdf[pages={1}]{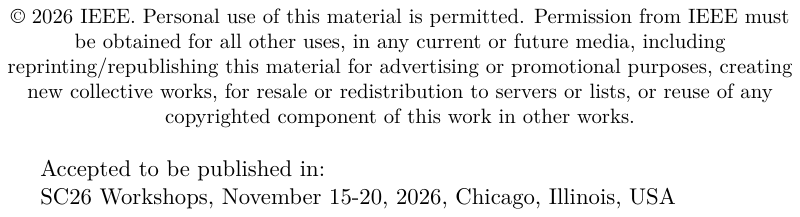}

\bstctlcite{IEEEexample:BSTcontrol} 

\title{The Shape of Speed: \\Impacts of Partition Geometry and Rank Density in Distributed Quantum Circuit Simulations}

\author{%
\IEEEauthorblockN{Yikai Mao\IEEEauthorrefmark{1},
Yuan He\IEEEauthorrefmark{1},
Shaowen Li\IEEEauthorrefmark{1},
Masaaki Kondo\IEEEauthorrefmark{1}\IEEEauthorrefmark{2}}
\IEEEauthorblockA{\IEEEauthorrefmark{1}RIKEN Center for Computational
Science, Kobe, Hyogo, Japan}
\IEEEauthorblockA{\IEEEauthorrefmark{2}Keio University, Yokohama, Kanagawa,
Japan\\
\{yikai.mao, yuan.he.uw, shaowen.li, masaaki.kondo\}@riken.jp}}


\maketitle
\thispagestyle{fancy}
\lhead{}
\rhead{}
\chead{}
\lfoot{\footnotesize{
SC26 Workshops, November 15-20, 2026, Chicago, Illinois, USA
\newline 979-8-3195-1221-5/26/\$31.00 \copyright 2026 IEEE}}
\rfoot{}
\cfoot{}
\renewcommand{\headrulewidth}{0pt}
\renewcommand{\footrulewidth}{0pt}

\begin{abstract}

In distributed quantum circuit simulation, a poorly shaped partition can halve performance before computation begins. Evaluation on Fugaku across 764 validated configurations (twelve algorithms, thirteen torus partition geometries, and six rank densities for 39-qubit simulations on 1,024 nodes) shows that partition geometry dominates runtime. All twelve algorithms run 1.73–2.31x slower on flat partitions than on near-cubic ones despite identical data transfer, proving the slowdown stems from network delivery rather than communication volume. This penalty scales with the 3D torus partition aspect ratio (runtime $\propto a^{0.39}$, $r = 0.72$). Rank density is secondary, cutting runtime by 11\% at 16 ranks per node only on compact geometries. Ultimately, requesting a near-cubic partition with 16 ranks per node roughly halves time-to-solution relative to flat partitions, which also consume 1.82x more energy. A simulator-free all-to-all microbenchmark confirms a similar geometry penalty for collective-dominated workloads.
\end{abstract}


\section{Introduction}
\label{sec:intro}

Classical simulation of quantum circuits has become a production
workload on HPC systems. These simulations validate algorithms before
quantum hardware execution, verify noisy device outputs against exact
\rev{simulated amplitudes}~\cite{villalonga2019flexible}, and \rev{provide noise-free reference results for circuits that current hardware cannot yet execute reliably}, making them a standing
service of emerging Quantum--HPC ecosystems.
At production scales, this computational workload introduces severe
resource demands: simulating a 39-qubit state vector requires 8\,TiB of memory,
occupying on the order of a thousand nodes once communication buffers
and practical per-node memory budgets are accounted for. \rev{At this scale}, simulation performance is predominantly
communication-bound: applying a general, non-diagonal gate to a globally
distributed qubit requires \rev{exchanging state-vector data between pairs of ranks}
across the interconnect~\cite{haner2017petabyte,imamura2022mpiqulacs}. Consequently,
execution time is constrained by network performance, which \rev{depends strongly on} allocation decisions made prior to job initiation.

Two primary parameters govern this behavior: the user-specified number of
Message Passing Interface (MPI) ranks allocated per node, and the scheduler-determined geometry of the
contiguous torus partition assigned to the job. Currently, the latter
parameter is rarely controlled. \rev{Eleven 1,024-node
simulation jobs submitted to Fugaku with identical node requests received} five distinct partition geometries;
eight of the eleven allocations were highly elongated. Because production job packers
typically optimize for machine utilization and standard submission scripts carry no
information regarding application communication structure, network-bound
workloads can silently lose \rev{about half their performance, and more on the flattest shapes,} between otherwise identical submissions.

This paper quantifies these performance variations under controlled
conditions. Achieving such experimental control is non-trivial: on Fugaku, the batch scheduler \rev{accepted explicit shape requests only at partition sizes that are multiples of 96 nodes (consistent with I/O alignment)} in our campaigns, and these sizes are never powers of two, whereas distributed state-vector
simulators require power-of-two rank counts. We resolve this constraint
through a partial-use mechanism: requesting an explicitly shaped partition
and executing on a power-of-two subset of the allocated nodes. Using this
technique, we evaluate twelve quantum algorithms across thirteen torus
geometries and six rank densities for 39-qubit simulations on 1,024 nodes,
producing \rev{764 validated configurations (936 runs minus 172 that received a partition of a different size and shape than requested).}

Our empirical measurements reveal a clear hierarchy between these two
control variables. Partition geometry is the primary performance driver:
across all algorithms, flat partitions execute \rev{1.73--2.31$\times$} slower
than near-cubic geometries. Because total injected network traffic remains
constant across shapes, geometry \rev{changes how quickly the network delivers that traffic, not how much traffic there is}. Rank density acts as a secondary, conditional factor: \rev{on compact partitions, allocating 16 ranks per node shortens average runtime by 11\% relative to one rank per node, but it} provides no measurable benefit on flat allocations. This
conditional dependency \rev{means that rank-density measurements taken without controlling placement can appear contradictory and non-monotonic.} \rev{The dominant geometry effect} can be modeled to first order using the partition's aspect ratio alone, providing a practical \rev{submission-time predictor of execution time and, because energy closely tracks time, of energy consumption}.
These results form an immediately deployable best practice \rev{whose reach likely extends beyond quantum simulation: a simulator-free all-to-all microbenchmark shows a similar geometry penalty, so other communication-bound workloads dominated by all-to-all or transpose-style collectives on contiguous torus partitions should benefit as well}. Distributed quantum circuit simulation is a demanding and timely instance: a single 39-qubit \rev{job} moves 0.3 to 3.2\,PB across
the interconnect, depending on the circuit.

Our primary contributions are as follows:
\begin{itemize}
\item To our knowledge, the first controlled characterization of
partition-geometry effects on distributed quantum circuit simulation on a
production supercomputer, including the allocation methodology that makes
such controlled experiments feasible (Sections~\ref{sec:proposal}
and~\ref{sec:eval}).
\item Empirical identification of the placement hierarchy: partition geometry serves as a $\sim$2$\times$ performance lever \rev{consistent across all twelve algorithms}, whereas rank density acts as a conditional lever \rev{(11\% shorter runtime)} dependent on compact geometry (Sections~\ref{sec:proposal:hierarchy}, \ref{sec:eval:shape},
and~\ref{sec:eval:interaction}).
\item Formulation of a single-parameter surrogate model ($\text{runtime}
\propto \aspect^{\rev{0.39}}$, $r = \rev{0.72}$) \rev{that gives workload managers an evaluable submission-time cost model, and of an associated best-practice execution strategy that improves time-to-solution by roughly 2$\times$ relative to flat partitions, which also consume 1.82$\times$ more energy}
(Sections~\ref{sec:proposal:surrogate}, \ref{sec:proposal:practice},
\ref{sec:eval:predictor}, and~\ref{sec:eval:energy}).
\end{itemize}

The paper is structured as follows:
Section~\ref{sec:background} reviews related work. Section~\ref{sec:proposal}
describes the placement hierarchy, recommended practices, and allocation
mechanism. Section~\ref{sec:eval} presents our empirical evaluation.
Section~\ref{sec:discussion} discusses limitations and system implications,
and Section~\ref{sec:conclusion} concludes.

\section{Background and Related Work}
\label{sec:background}

\subsection{Distributed State-Vector Simulation and Its Communication}

State-vector simulation stores all $2^{n}$ complex amplitudes of an
$n$-qubit register and applies gates as sparse linear operators over
them~\cite{jones2019quest,suzuki2021qulacs,guerreschi2020intel}. \rev{Beyond roughly 30 qubits (16\,GiB) on a 32\,GiB Fugaku node, the state vector exceeds a single node's memory} and must be distributed across multiple nodes. \rev{The qubits are consequently partitioned into rank-local and global qubits; a non-diagonal gate on a global qubit pairs each rank with the partner whose index differs in that qubit's bit, and the two exchange state-vector data, over the interconnect whenever the partners reside on different nodes}~\cite{haner2017petabyte,imamura2022mpiqulacs}.
\rev{Large-scale state-vector simulations on \rev{petascale} systems~\cite{haner2017petabyte,deraedt2019massively} and \rev{multi-GPU simulators built on the cuQuantum SDK}~\cite{cuquantum2023} report that performance at scale
is limited largely by data movement rather than arithmetic, whereas tensor-network simulators~\cite{pednault2017breaking,villalonga2019flexible} avoid much of this communication at the cost of other trade-offs.} An extensive body of
literature addresses this bottleneck from the application layer by reducing \rev{data movement or memory footprint} via gate scheduling, cache blocking, and gate
fusion~\cite{cacheblocking2026}, qubit
reordering~\cite{lazyqr2024}, data compression~\cite{wu2020full}, and
communication-avoiding \rev{partitioning}~\cite{chen2018high}. Recent benchmarks on GPU
clusters also identify \rev{interconnect performance as a key scaling
limit}~\cite{multigpunet2025}. While prior \rev{simulator} work focuses on optimizing the
traffic generated by the application, it leaves open the question of how efficiently
the underlying network \emph{delivers} that traffic. \rev{On a torus interconnect that gives each job a contiguous, private partition, this delivery efficiency depends strongly on the partition's geometry}, which is fixed before computation begins.

\subsection{Topology-Aware Placement in HPC Systems}

The impact of job placement on performance is well established for classical
workloads across various network topologies. \rev{On Cray systems, Bhatele et al.\ observed run-to-run performance variations exceeding 30\%, caused mainly by network contention from neighboring jobs, whereas Blue Gene systems, which give each job an isolated torus partition, showed little variability~\cite{bhatele2013neighborhood}.} Consequently, scheduler research has
emphasized topology-aware allocation algorithms \rev{that minimize fragmentation while giving communication-sensitive jobs compact, convex shapes}~\cite{li2017topology}. This placement sensitivity is also observed on other interconnects:
job-interference studies on dragonfly networks trade locality against hotspot
avoidance~\cite{yang2016bully}, \rev{other studies link performance predictability
to allocation isolation~\cite{jokanovic2015quiet}, and measurements on a modern
dragonfly system show that hardware congestion control can largely mitigate such interference~\cite{desensi2020slingshot}.}
What distinguishes \rev{systems that allocate contiguous, private torus partitions} is that the placement variable is explicit
and low-dimensional. \rev{On Fugaku, each large job receives a contiguous block of the Tofu Interconnect D network, which it sees through Tofu's virtual three-dimensional torus rank mapping~\cite{ajima2018tofud}.} While this contiguous allocation \rev{largely} eliminates interference from neighboring jobs, it leaves the partition's own geometry as a dominant,
yet largely unexamined, performance factor. 
\rev{Simulation studies of fat-tree clusters found that the cost of contiguous allocation is compensated when job runtimes improve by roughly 20--30\%~\cite{pascual2009job}, and work on IBM Blue Gene/Q showed that relaxing network allocation constraints improves scheduling, especially when it accounts for each job's communication sensitivity~\cite{zhou2016improving}. Oltchik and Schwartz further showed that the shape of an allocated partition can cause avoidable contention and derived a predictor from partition structure~\cite{oltchik2020network}; our measurements complement theirs with \rev{quantum-simulation workloads on Fugaku's Tofu~D network}.} 
\rev{This topology sensitivity has re-emerged prominently in modern accelerated architectures: for example, Slurm's Block Topology plugin for NVIDIA GB200 NVL72 systems enforces contiguous placement to restrict all-to-all communication to high-bandwidth NVLink domains rather than crossing standard network fabrics~\cite{nvidia2024slurm}.}
\rev{Apart from these few studies,} prior topology-aware research has primarily optimized rank-to-node mappings \emph{within} a fixed allocation or evaluated allocation
heuristics via simulation. Controlled, application-level measurements of allocation
geometry on production systems remain scarce, particularly for distributed quantum
circuit simulation, where \rev{every non-diagonal global-qubit gate exchanges large state-vector slices between pairs of ranks that may lie far apart in the partition, a severe stress test} for network placement.

\subsection{Benchmarking and Workload Management}

Standardized benchmark suites and circuit
generators~\cite{mao2025qgen,tomesh2022supermarq,li2023qasmbench} define
\emph{what} to run when evaluating quantum circuit simulation, and published
evaluations report scaling with qubit count, node count, and rank
configuration. What common benchmarking practice rarely controls, or even
records, is the geometry of the allocation each measurement ran on.
\rev{Our controlled grid shows that geometry alone changes runtime by about 2$\times$ at identical traffic (Section~\ref{sec:eval:shape}) and determines whether tuning rank density pays off (Section~\ref{sec:eval:interaction}),} so
placement belongs in the measurement protocol of any communication-bound
benchmark, alongside the variability controls motivated in the placement
literature above. Systems that schedule quantum workloads alongside
classical ones~\cite{pilotquantum2024,qpusharing2026,elephantscheduling2025,qrmi2026}
\rev{focus on sharing quantum resources and generally leave the network placement of classical jobs to the underlying batch scheduler, which yields \rev{default allocations like those in Section~\ref{sec:proposal:motivation}}.} The aspect-ratio surrogate of
Section~\ref{sec:proposal:surrogate} addresses this gap with an empirical,
single-parameter cost model evaluable at job submission, and the practice of
Section~\ref{sec:proposal:practice} makes the benefit available to users
without any scheduler changes.

\section{From Placement Lottery to Best Practice}
\label{sec:proposal}

\begin{figure}[t]
\centering
\includegraphics[width=\columnwidth]{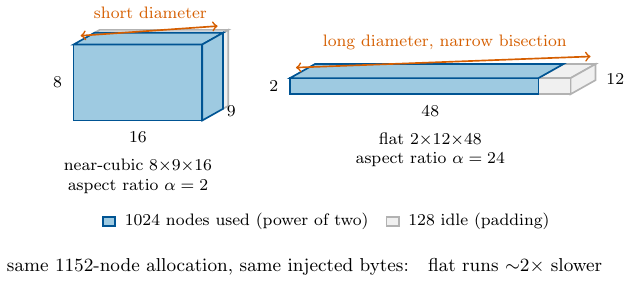}
\caption{Two possible geometries for a 1,152-node Tofu allocation, \rev{drawn roughly to scale, with thin axes widened and the long axis shortened for legibility}. The simulation runs on a power-of-two subset of 1,024 nodes
(blue); 128 nodes idle as padding (gray), the same 12.5\% overhead that
the scheduler's automatic padding imposes on unshaped requests. Both partitions carry identical traffic, but the flat one drains it through a longer diameter and a narrower bisection. \rev{Averaged over all flat geometries ($\aspect \ge 16$), runs take about twice as long as on near-cubic ones (the $\sim$2$\times$ in the figure); the drawn pair differs by 3.2$\times$.}}
\label{fig:schematic}
\end{figure}

\begin{table}[t]
\centering
\caption{Geometries assigned by default placement across eleven
1,024-node submissions \rev{with identical node requests} on Fugaku; the scheduler padded each request to the
1,152-node partitions listed. The aspect ratio $\aspect$ is the longest
axis divided by the shortest.}
\label{tab:default}
\begin{tabular}{@{}lrr@{}}
\toprule
Assigned geometry & Aspect ratio \aspect & Frequency \\
\midrule
$2{\times}12{\times}48$ & 24 (flattest) & 5 \\
$2{\times}18{\times}32$ & 16 & 3 \\
$4{\times}9{\times}32$  & 8  & 1 \\
$8{\times}6{\times}24$  & 4  & 1 \\
$8{\times}9{\times}16$  & 2 (near-cubic) & 1 \\
\bottomrule
\end{tabular}
\end{table}

\subsection{The Placement Lottery of Default Allocation}
\label{sec:proposal:motivation}

\rev{When the node count is set by memory, that is, by the state vector and its communication buffers rather than by computation}, distributed state-vector simulation is fundamentally limited by communication
(Section~\ref{sec:background}). At 39 qubits, the
8\,TiB state vector \rev{is distributed over 1,024 Fugaku nodes (8\,GiB per node), \rev{leaving room for communication buffers}}. Under these
operating conditions, execution time is \rev{governed} by two parameters fixed \rev{before execution begins}: the geometry of the contiguous torus partition assigned to the job
(Figure~\ref{fig:schematic}), and the number of MPI ranks allocated per node.

Currently, users rarely specify or control partition geometry. 
\rev{We characterize geometry using the partition aspect ratio, defined as $\aspect = \max(A,B,C)/\min(A,B,C)$ for a contiguous sub-torus of dimensions $A \times B \times C$, where $\aspect = 1$ denotes a perfect cube and larger values indicate increasingly elongated allocations.} Table~\ref{tab:default}
reports the geometries assigned to eleven 1,024-node submissions \rev{(different circuits and rank densities)} where
only the total node count was requested. The scheduler padded each request to a
1,152-node contiguous partition and assigned five different geometries\rev{, presumably} depending on free-space availability across the torus. Flat allocations ($\aspect \ge 16$) account
for eight of the eleven submissions, with the flattest geometry ($\aspect = 24$)
assigned five times. This pattern is consistent with packing policies that preserve compact
free regions while filling elongated remnants. Consequently,
default placement is neither reproducible nor performance-aware. Because \rev{flat partitions ($\aspect \ge 16$)} execute this workload roughly twice as slowly as \rev{near-cubic} geometries (as shown in Section~\ref{sec:eval}), standard job submissions can forfeit \rev{about half of their potential performance, and more on the flattest geometries}. Moreover, standard submission interfaces do not capture an application's
communication characteristics, so schedulers cannot optimize for this
behavior automatically.

The remainder of this section analyzes data from \rev{764} controlled configurations
(Section~\ref{sec:eval}) to establish a two-level placement hierarchy
(Section~\ref{sec:proposal:hierarchy}), propose a scalar performance predictor
(Section~\ref{sec:proposal:surrogate}), outline recommended user practices
(Section~\ref{sec:proposal:practice}), and detail the allocation mechanism that
enables controlled geometry experiments on an unmodified production scheduler
(Section~\ref{sec:proposal:mechanism}).

\subsection{Geometry Dominates and Constrains Rank Density}
\label{sec:proposal:hierarchy}

Partition geometry is the primary performance determinant. Across all twelve
algorithms evaluated, flat partitions ($\aspect \ge 16$) increase execution time by
\rev{1.73--2.31$\times$ (mean 2.04$\times$)} compared to near-cubic partitions
($\aspect \le 3$, termed compact or cubic below), consistently across gate counts ranging from 91 to 1,404. \rev{This slowdown does not come from extra traffic: for a given algorithm and rank density, the injected data volume is the same on every geometry (Section~\ref{sec:eval:shape}), because the exchange pattern depends on the amplitude distribution, not on the physical mapping of ranks.} Partition geometry impacts performance
by modifying the partition diameter and bisection bandwidth, changing how
quickly the network drains those bytes, as Figure~\ref{fig:schematic}
illustrates. Because the bottleneck lies in the network rather than in any
circuit-specific computation, all twelve algorithms suffer nearly the
same slowdown.

Rank density acts as a secondary factor whose effectiveness is \emph{conditional
on geometry}. On near-cubic partitions, allocating 16 ranks per node (3 OpenMP
threads per rank) yields an average \rev{11\%} \rev{shorter runtime than one rank per node}. Conversely, on flat
partitions, rank-density adjustments yield no measurable benefit \rev{(within 2\%, Table~\ref{tab:interaction}), consistent with execution being bounded by inter-node network delivery, which changes in rank density cannot alleviate}. Oversubscription (32 ranks $\times$ 2 threads $= 64$ threads on
48 cores) \rev{performs worse on average than 16 ranks per node}, although two algorithms (QAOA, QW) achieve
optimal runtimes under this configuration (Table~\ref{tab:shape}); as a general
default, oversubscription should be avoided.

This conditional relationship \rev{can make} rank-density tuning \rev{appear}
inconsistent wherever
placement goes uncontrolled\rev{, as in our own earlier campaign (Figure~\ref{fig:interaction}), where node count also varied}: under default scheduler
policies, partition geometry varies arbitrarily between runs. Consequently, measurements across different submissions
conflate evaluations on compact geometries (where density impacts performance) with
evaluations on flat geometries (where it does not). Our controlled grid isolates these variables.

\subsection{Aspect Ratio as a Predictive Surrogate}
\label{sec:proposal:surrogate}

Among candidate geometric metrics, the partition aspect ratio, 
\rev{introduced in Section~\ref{sec:proposal:motivation},} is the strongest predictor of
execution time. After normalizing scale differences across algorithms and densities, the
correlation between log runtime and $\log \aspect$ across all \rev{764} configurations
is $r = \rev{0.72}$, outperforming inverse bisection width \rev{(the reciprocal of the product of the two shorter axes)} ($r = \rev{0.67}$) and inverse shortest
axis length ($r = \rev{0.61}$). A power-law regression across the dataset yields:
\begin{equation}
T(\aspect) \;\approx\; T_{1} \cdot \aspect^{\,\rev{0.39}},
\label{eq:surrogate}
\end{equation}
where $T_{1}$ \rev{is the runtime extrapolated to $\aspect = 1$ (a perfect cube, not realizable at this node count)} for a given algorithm and density. Eq.~\eqref{eq:surrogate} predicts a $\rev{2.6}\times$ execution-time penalty when shifting from $\aspect = 2$ to $\aspect = 24$ \rev{and a 2.2$\times$ ratio between the near-cubic and flat classes, somewhat above the observed 2.04$\times$; individual geometries depart from the fit (Section~\ref{sec:eval:predictor})}.
Partition elongation predicts performance degradation somewhat better
than minimum-cut bandwidth alone. \rev{Because these metrics are strongly correlated across the 13 geometries ($r = 0.87$--$0.97$), and a mean-hop-distance proxy (the sum of the axes) scores similarly ($r = 0.70$), the data cannot separate path length from bisection bandwidth.} Application-level
metrics carry comparatively little predictive weight (gate count vs.\ topology
sensitivity: $r = \rev{-0.27}$), allowing this surrogate model to function without
per-application calibration beyond the scaling constant $T_{1}$.

Because this surrogate metric can be evaluated at job submission using only partition
dimensions, it provides a \rev{simple cost model for distributed state-vector simulation on Fugaku}. 
\rev{While an empirical correlation of $r = 0.72$ indicates moderate scatter around the regression line\rev{, including geometry-specific departures} (discussed in Section~\ref{sec:eval:predictor}), Eq.~\eqref{eq:surrogate} serves as a lightweight, submission-time screening heuristic rather than an exact analytical model.} Given a set
of available torus regions, a scheduler could score feasible placements by $\aspect^{\rev{0.39}}$,
weighing the predicted slowdown of an elongated allocation against the wait time for a
compact region to co-optimize time-to-solution and energy consumption. \rev{Flat partitions also consume 1.82$\times$ more energy than near-cubic ones (Section~\ref{sec:eval:energy}), so compact placements improve both performance and energy efficiency.} While building an autonomous allocator
is outside the scope of this paper, our measurements provide the empirical cost model
needed to support one.

\subsection{Recommended Execution Practice}
\label{sec:proposal:practice}

For end users, these empirical findings suggest a straightforward operational procedure:

\begin{enumerate}
\item \textbf{Request geometry explicitly:} Specify an I/O-aligned, near-cubic
partition ($\aspect \le 3$\rev{; in our study, $8{\times}9{\times}16$ or $8{\times}18{\times}8$, as $6{\times}12{\times}16$ ran markedly slower}) and execute using power-of-two partial node utilization
(Section~\ref{sec:proposal:mechanism}).
\item \textbf{Optimize rank density:} Allocate 16 ranks per node as the standard
baseline; \rev{tune density only when both the workload and the geometry are known}
(Section~\ref{sec:eval:interaction}).
\item \textbf{Handle flat allocations:} If only flat partitions are available,
rank-density adjustments will not improve performance. Users should either defer
the job until a compact region is available (using Eq.~\eqref{eq:surrogate} to
quantify the cost of immediate execution) or \rev{run immediately at any density (e.g., $R = 16$)}.
\end{enumerate}

Adopting these practices improves time-to-solution by roughly 2$\times$ \rev{relative to flat partitions, which default placement assigned to eight of eleven submissions and which also consume 1.82$\times$ more energy than near-cubic ones} (Section~\ref{sec:eval}). This approach
requires no modifications to the simulation software or underlying scheduler and can be applied \rev{today on Fugaku and on other systems whose schedulers accept explicit partition shapes, provided the allocated shape is checked (Section~\ref{sec:proposal:mechanism})}.

\subsection{Obtaining Defined Geometries on Production Schedulers}
\label{sec:proposal:mechanism}

Conducting controlled geometry experiments on a production system requires overcoming
a structural scheduling constraint. On Fugaku, a standard 1,024-node submission is
padded automatically to a larger contiguous partition of scheduler-determined
geometry, 1,152 nodes in all eleven default submissions of our campaign
(occasionally other sizes, such as a 1,200-node case observed in the
microbenchmark campaign of Section~\ref{sec:eval:micro},
Table~\ref{tab:micro}). Requests for explicit shapes at unaligned counts (e.g., $16{\times}8{\times}8 = 1{,}024$)
are overridden by scheduler padding rules. In our campaigns, explicit geometries were honored only when the
requested node count was a multiple of 96 ($1{,}152 = 12 \times 96$),
consistent with the system's I/O-alignment granularity. As a result, alignable counts such as $1{,}152 = 2^{7}\,3^{2}$
are never powers of two. However, distributed state-vector simulators require a
power-of-two rank count to partition $2^{n}$ amplitudes evenly. \rev{The two constraints therefore have no common solution: no aligned node count, fully used at a uniform number of ranks per node, yields a power-of-two rank count, and neither constraint can be relaxed.} This incompatibility helps explain why controlled placement studies of
this workload have been impractical \rev{on Fugaku} to date.

We resolve this incompatibility through \emph{partial use}: requesting an explicitly
shaped, I/O-aligned partition (1,152 nodes) and executing the simulation on a power-of-two
subset ($1{,}024 \cdot R$ ranks across 1,024 nodes at $R$ ranks per node), leaving the remaining
128 nodes idle. This allows execution within a controlled, user-specified geometry at the
cost of a 12.5\% idle-node allocation overhead. Crucially, this overhead matches the padding cost that \rev{default placement imposed on all eleven unshaped submissions of Table~\ref{tab:default}}; partial use thus
converts unmanaged padding into an experimental control variable. 
\rev{Even against a hypothetical zero-padding 1,024-node allocation that lands on a flat partition, the 2.04$\times$ speedup outweighs the idle nodes, yielding a net advantage of $(1{,}024/1{,}152) \times 2.04 \approx 1.8\times$ in completed simulation work per allocated node-hour.} \rev{We requested thirteen distinct 1,152-node geometries, each of which the scheduler granted in at least some runs,} spanning aspect ratios from $\aspect = 2$ (near-cubic $8{\times}9{\times}16$) to $\aspect = 24$ (flat $2{\times}12{\times}48$)\rev{; these provide} the experimental basis for Section~\ref{sec:eval}. \rev{The scheduler did not always honor these requests, however: 172 of our 936 runs (18\%) received a partition of a different size and shape and were excluded from all analyses, and 80 runs received the requested dimensions in a different axis order and were retained as separate orientation variants (e.g., $16{\times}9{\times}8$ for a requested $8{\times}9{\times}16$).}

\section{Evaluation}
\label{sec:eval}

This section details the experimental methodology supporting
Section~\ref{sec:proposal} and presents the full evidence: the setup
(IV-A), the geometry effect (IV-B), the shape--density interaction
(IV-C), the predictor comparison (IV-D), energy scaling (IV-E), and a
simulator-free microbenchmark (IV-F).

\subsection{Experimental Setup}
\label{sec:eval:setup}

\textbf{Platform.} Experiments were conducted on Fugaku, which features 48-core
Fujitsu A64FX processors with 32\,GiB of HBM2 memory (1,024\,GB/s bandwidth), connected
by the Tofu Interconnect D 6D mesh/torus network~\cite{ajima2018tofud} providing ten 6.8\,GB/s links per
node. Evaluations used the MPI-parallel implementation of Qulacs
for A64FX~\cite{imamura2022mpiqulacs}. Hardware counters provided measurements for node
power consumption (to calculate total energy) and Tofu network user traffic (to measure
communication volume).

\textbf{Workload.} To avoid confounding geometric sensitivity with problem size, every
configuration simulated $n = 39$ qubits. The resulting 8\,TiB state vector allocates
8\,GiB per node across 1,024 nodes, \rev{a node count set by memory, as in production distributed quantum simulation}. The evaluation suite comprises twelve algorithms spanning the query,
communication, variational, Fourier, and search categories
\rev{of the Q-Gen generator}~\cite{mao2025qgen} (Table~\ref{tab:workload}). Because standard
benchmarking suites do not provide these circuits at 39 qubits, each circuit was generated
programmatically at full width \rev{as a representative fixed-depth kernel of each algorithm rather than a complete textbook implementation (e.g., QPE omits the final inverse QFT, and the quantum-counting kernel is a phase-estimation circuit)}. For high-depth algorithms
(\rev{Grover, QC, QPE, QW, Shor}), iteration counts were fixed at small constants to ensure jobs completed
within scheduler wall-time limits. Gate counts range from 91 (QKD) to 1,404 (VQC), with
every circuit operating on the full 39-qubit register. 
\rev{At 39 qubits, execution time is largely determined by global state-vector exchanges over the network. Because production runs of iterative algorithms (e.g., repeated VQE or QAOA ansatz evaluations) repeat these same exchanges, the geometry penalty measured on one kernel pass is expected to carry over to end-to-end execution time roughly proportionally. For example, the mean VQC pass drops from 1,576\,s on flat partitions to 912\,s on near-cubic ones (Table~\ref{tab:shape}), saving about 213 node-hours per pass on a 1,152-node allocation; an iterative run repeats this saving at every pass, and obtaining it requires only an explicit, verified shape in the job request (Section~\ref{sec:proposal:mechanism}).}

\begin{table}[t]
\centering
\caption{Workload: twelve quantum algorithms, all at $n=39$ qubits (8\,TiB state,
8\,GiB per node on 1,024 nodes), sorted by category and algorithm name.
\rev{Categories follow Q-Gen~\cite{mao2025qgen}. All entries are fixed-depth representative kernels; e.g., VQE is a single ansatz evaluation, and the Shor-style kernel uses CNOT patterns and a truncated QFT in place of modular-exponentiation arithmetic.}}
\label{tab:workload}
\small
\begin{tabular}{@{}llr@{}}
\toprule
Category & Algorithm & Gates \\
\midrule
\multirow{2}{*}{Communication} & Quantum Key Distribution (QKD) & 91 \\
 & Quantum Teleportation (QT) & 114 \\
\midrule
\multirow{2}{*}{Fourier} & Quantum Phase Estimation (QPE) & 154 \\
 & \rev{Shor-style kernel (Shor)} & \rev{356} \\
\midrule
\multirow{2}{*}{Query} & Bernstein--Vazirani (BV) & 96 \\
 & Deutsch--Jozsa (DJ) & 115 \\
\midrule
\multirow{3}{*}{Search} & Grover's Algorithm (Grover) & 735 \\
 & \rev{Quantum Counting (QC)} & \rev{192} \\
 & Quantum Walk (QW) & 770 \\
\midrule
\multirow{3}{*}{Variational} & Quantum Approx.\ Optim.\ (QAOA) & 507 \\
 & Var.\ Classifier (VQC) & 1,404 \\
 & Var.\ Quantum Eigensolver (VQE) & 462 \\
\bottomrule
\end{tabular}
\end{table}

\textbf{Sweep design.} The factorial experimental grid comprises 12 algorithms $\times$
13 geometries $\times$ 6 rank densities $= 936$ configurations. Each configuration was
executed as an independent batch job using the explicit placement mechanism described in
Section~\ref{sec:proposal:mechanism}, ensuring that geometry and rank density remained unconfounded
and that hardware counters reflected individual configuration performance. Evaluated rank densities
included $R \in \{1, 2, 4, 8, 16, 32\}$, using $48/R$ OpenMP threads per rank for $R \le 16$
(close binding, all 48 cores occupied) and 2 threads per rank at $R = 32$, representing the sole
oversubscribed configuration (64 threads on 48 cores). \rev{The power-of-two rank constraint excluded $R = 24$ and $R = 48$, and the scheduler limit of 48 processes per node excluded $R = 64$.} MPI ranks used the scheduler's default rank-to-node mapping in every
cell, with no explicit rank-map files supplied, so the mapping policy was
uniform across the grid. During execution, each job constructed the circuit, performed one \rev{warm-up} pass, and recorded the execution time of one full-circuit update pass; across the grid, the warm-up pass agreed with the timed pass to a median difference of \rev{0.6\%}.
\rev{A run counted as completed only if its timed pass finished and its result record confirmed $n = 39$ and exactly 1,024 distinct execution nodes. Failed or timed-out jobs were resubmitted until every cell had a completed run.} \rev{For each cell, the latest completed run was retained only if the scheduler record showed that its allocated partition matched the requested dimensions, in any axis order; 172 of the 936 cells (18\%) failed this check and were excluded, leaving 764 runs on which all application-grid results are based. } Alongside the application grid, a pure MPI all-to-all
microbenchmark provided a simulator-free evaluation of geometry effects
(setup and results in Section~\ref{sec:eval:micro}).

\subsection{Partition Geometry Dominates Performance}
\label{sec:eval:shape}

Table~\ref{tab:shape} quantifies the primary level of the placement hierarchy. For every
evaluated algorithm, the fastest execution time across \rev{its retained} (geometry, density) combinations
occurred on a near-cubic geometry: the \rev{Best cell} column contains only \rev{near-cubic partitions: $8{\times}9{\times}16$ or its rotated orientation $16{\times}9{\times}8$} (eight algorithms, typically at $R=16$) and $8{\times}18{\times}8$ at $R=1$ (the remaining four). The ratio of mean execution time on flat partitions to
near-cubic partitions falls within a \rev{band of 1.73--2.31$\times$ (mean 2.04$\times$)},
despite best-configuration runtimes spanning 52\,s (QT) to 709\,s (QC) and gate counts ranging from
91 to 1,404.

\begin{table}[t]
\centering
\caption{Per-algorithm results, sorted by algorithm name\rev{, over retained runs}. Best cell over all \rev{retained} (geometry, rank density $R$) configurations; class means average over \rev{all retained (geometry, $R$) cells of the class} (cubic: $\aspect \le 3$; flat: $\aspect \ge 16$\rev{; 6--18 cells per class and algorithm}); flat/cubic ratios for execution time
and energy to solution. Ratios are computed from unrounded means.}
\label{tab:shape}
\footnotesize
\setlength{\tabcolsep}{3.5pt}
\begin{tabular}{@{}lrlrrrr@{}}
\toprule
 & \multicolumn{2}{c}{Best cell} & \multicolumn{2}{c}{Mean (s)} & \multicolumn{2}{c}{Flat/cubic} \\
\cmidrule(lr){2-3}\cmidrule(lr){4-5}\cmidrule(l){6-7}
Algorithm & (s) & geometry, $R$ & Cubic & Flat & Time & Energy \\
\midrule
BV     & \rev{128} & $8{\times}18{\times}8$, 1 & 216 & \rev{482} & \rev{2.23$\times$} & \rev{1.95$\times$} \\
DJ     & 193 & $8{\times}18{\times}8$, 1 & 357 & \rev{750} & \rev{2.10$\times$} & \rev{1.95$\times$} \\
Grover & 413 & $8{\times}9{\times}16$, 16 & 571 & \rev{1,221} & \rev{2.14$\times$} & \rev{1.92$\times$} \\
QAOA   & \rev{236} & \rev{$16{\times}9{\times}8$}, 32 & \rev{333} & \rev{710} & \rev{2.13$\times$} & \rev{1.93$\times$} \\
QC     & 709 & $8{\times}18{\times}8$, 1 & \rev{1,411} & \rev{2,911} & \rev{2.06$\times$} & \rev{1.90$\times$} \\
QKD    & 59 & \rev{$16{\times}9{\times}8$}, 16 & 85 & \rev{184} & \rev{2.16$\times$} & \rev{1.69$\times$} \\
QPE    & 497 & $8{\times}18{\times}8$, 1 & \rev{1,044} & \rev{2,004} & \rev{1.92$\times$} & \rev{1.82$\times$} \\
QT     & 52 & \rev{$16{\times}9{\times}8$}, 16 & 70 & \rev{131} & \rev{1.88$\times$} & \rev{1.57$\times$} \\
QW     & 300 & $8{\times}9{\times}16$, 32 & \rev{404} & \rev{935} & \rev{2.31$\times$} & \rev{2.05$\times$} \\
Shor   & 208 & $8{\times}9{\times}16$, 16 & 284 & 516 & 1.82$\times$ & 1.69$\times$ \\
VQC    & 660 & $8{\times}9{\times}16$, 16 & \rev{912} & \rev{1,576} & \rev{1.73$\times$} & \rev{1.62$\times$} \\
VQE    & 293 & \rev{$16{\times}9{\times}8$}, 16 & \rev{430} & \rev{840} & \rev{1.95$\times$} & \rev{1.73$\times$} \\
\midrule
Mean   &     &                            &      &      & \rev{2.04$\times$} & \rev{1.82$\times$} \\
\bottomrule
\end{tabular}
\end{table}

\begin{figure}[t]
\centering
\includegraphics[width=\columnwidth]{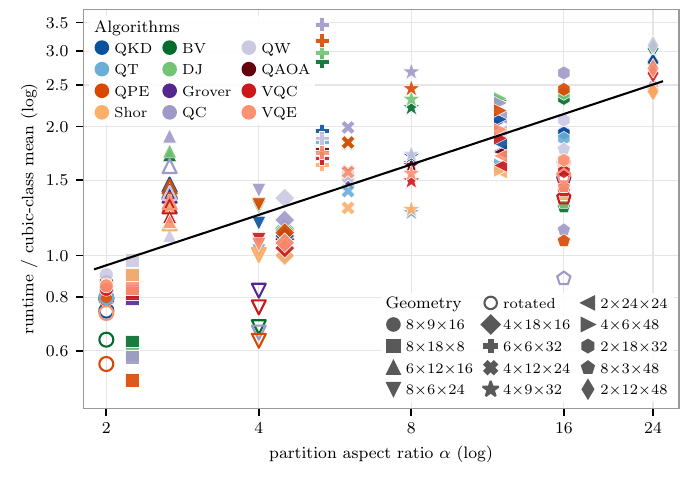}
\caption{Runtime versus partition aspect ratio, \rev{the 173 (algorithm, allocated partition) pairs with retained runs} (density-averaged, normalized to each algorithm's mean over near-cubic \rev{cells}, $\aspect \le 3$), with the fitted power law
$T \propto \aspect^{\rev{0.39}}$. \rev{Color identifies the algorithm (one color family per category of Table~\ref{tab:workload}) and marker shape the geometry; hollow markers show runs that received the requested dimensions in a different axis order. The 13 geometries yield 11 distinct aspect ratios because two pairs coincide: $\aspect = 12$ ($2{\times}24{\times}24$ and $4{\times}6{\times}48$) and $\aspect = 16$ ($2{\times}18{\times}32$ and $8{\times}3{\times}48$). Most algorithms follow the upward trend, with geometry-specific departures discussed in Section~\ref{sec:eval:predictor}.}}
\label{fig:aspect}
\end{figure}

Figure~\ref{fig:aspect} illustrates this relationship continuously across aspect ratios.
When aggregated across all algorithms, normalized runtime trends upward
with aspect ratio, following the fitted power law of
Section~\ref{sec:proposal:surrogate} with substantial per-geometry
scatter ($r = \rev{0.72}$).

\rev{These differences arise from how the network delivers the traffic, not from its volume.} For
a given algorithm, total user data injected into the Tofu network remains invariant
across partition geometries: per-job interconnect counters confirm that, at 16 ranks
per node, DJ injects 668.5\,TB and VQC injects 3,166.7\,TB, remaining constant across
\rev{all retained geometries (13 for DJ, 12 for VQC)} to within $10^{-6}$\,\%. Because communication volume is fixed
by the amplitude distribution, elongated geometries must transfer identical byte volumes
across longer diameters and narrower bisections, requiring roughly twice the execution time.

We can now \rev{assess} the performance impact of default placement (Table~\ref{tab:default}).
\rev{Default placement assigned flat geometries ($\aspect \ge 16$) to eight of the eleven submissions, which run 2.04$\times$ slower than near-cubic ones on average. A default submission therefore typically takes roughly twice as long as the recommended execution practice (Section~\ref{sec:proposal:practice}),}
a penalty that rank-density tuning alone cannot overcome.

\subsection{Interaction Between Shape and Rank Density}
\label{sec:eval:interaction}

Table~\ref{tab:interaction} quantifies the secondary level of the placement hierarchy.
Because each geometry's \rev{retained} density runtimes are expressed relative to their own
internal mean, the geometry effect is divided out\rev{, so the residual variation reflects rank density; restricting to the (algorithm, geometry) pairs with all six densities retained changes no entry by more than 0.02}.

\begin{table}[t]
\centering
\caption{Interaction of shape class and rank density $R$: runtimes normalized to each \rev{(algorithm, geometry) pair's own mean over its retained $R$} (lower is better), averaged over \rev{the algorithms and geometries of each class}. Bold marks the best cell in the table.}
\label{tab:interaction}
\begin{tabular}{@{}lcccccc@{}}
\toprule
Shape class & $R{=}1$ & 2 & 4 & 8 & 16 & 32 \\
\midrule
Near-cubic ($\aspect \le 3$) & \rev{1.06} & \rev{1.03} & 1.01 & \rev{0.99} & \textbf{\rev{0.95}} & \rev{0.97} \\
Flat ($\aspect \ge 16$)      & \rev{0.99} & \rev{1.00} & \rev{0.99} & \rev{1.00} & \rev{1.01} & \rev{1.01} \\
\bottomrule
\end{tabular}
\end{table}

\begin{figure}[t]
\centering
\includegraphics[width=\columnwidth]{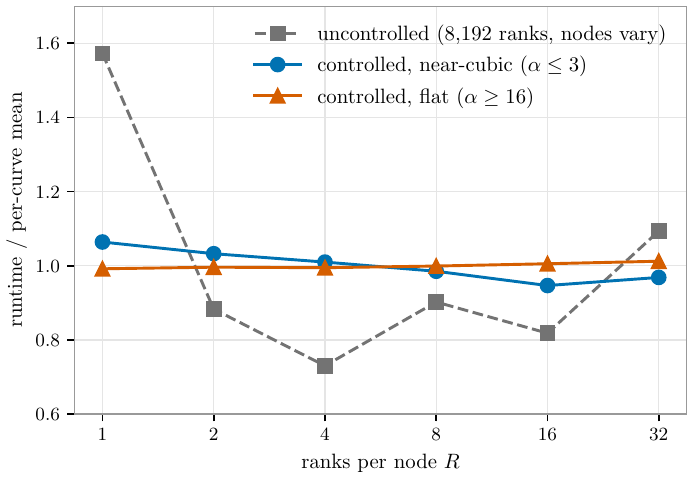}
\caption{Rank-density curves under the controlled protocol (near-cubic
vs.\ flat classes) and under the conventional uncontrolled protocol of
our earlier 15-algorithm campaign (total ranks fixed at 8,192, so node
count and placement vary with density); all three curves are normalized
to their own means. The uncontrolled curve swings by $2.15\times$ and
reverses direction\rev{, but it confounds placement with node count; the controlled curves isolate the conditional
effect}.}
\label{fig:interaction}
\end{figure}

On near-cubic geometries, rank density \rev{has a clear effect}: shifting from $R = 1$ to $R = 16$ reduces mean execution time by \rev{11\% (1.06 vs.\ the bolded 0.95)}, while \rev{the oversubscribed $R = 32$ is slower than $R = 16$}. Conversely, on flat geometries, the performance curve
remains essentially level \rev{(within 2\%)}.
\rev{This asymmetry is the conditional behavior summarized in Section~\ref{sec:proposal:hierarchy}; a plausible explanation is that rank density mainly changes intra-node execution, whose gains are visible on compact partitions but masked by longer network delivery times on flat ones.} \rev{Aggregated means also mask cell-level heterogeneity: on $8{\times}18{\times}8$, the fastest retained cell for BV, DJ, QC, and QPE is at $R = 1$ (Table~\ref{tab:shape}), whereas on $8{\times}9{\times}16$ $R = 16$ beats $R = 1$ for every algorithm with both runs. Therefore, $R = 16$ serves as the best general default, and density is worth tuning only when both the workload and the target geometry are known.}
Figure~\ref{fig:interaction} sets these controlled curves against our earlier
uncontrolled campaign, in which placement\rev{, node count, and resource group varied with density}.
Under that protocol, runtime \rev{swings} by more than twofold and \rev{the apparent density trend reverses};
controlling geometry \rev{and node count} isolates the
\rev{11\%} effect on compact partitions.

\subsection{Predictive Power of Aspect Ratio}
\label{sec:eval:predictor}

After normalizing scale across algorithms and densities, the correlation between log
runtime and $\log \aspect$ across all \rev{764} cells is $r = \rev{0.72}$, compared to $r = \rev{0.67}$
for inverse bisection width and $r = \rev{0.61}$ for inverse shortest axis length. The fitted
exponent of $\rev{0.39}$ (Eq.~\eqref{eq:surrogate}) predicts a $\rev{2.6}\times$ spread between the $\aspect = 2$ and
$\aspect = 24$ endpoints \rev{and a 2.2$\times$ ratio between the near-cubic and flat classes, against an observed 2.04$\times$; the measured endpoint pair differs by 3.2$\times$, reflecting the geometry-specific departures discussed below.} Circuit-level features fail to predict topology sensitivity reliably:
correlation with gate count is \rev{weak} and negative ($r = \rev{-0.27}$), and
individual gate-heavy workloads can still show high shape sensitivity
(e.g., \rev{QW at 770 gates shows a 2.31$\times$ ratio}), because gate
count does not reflect the proportion of global-qubit communication.

\rev{The power law captures the dominant trend, but individual geometries depart from it (Figure~\ref{fig:aspect}). Averaged over algorithms, $6{\times}12{\times}16$ ($\aspect \approx 2.67$) runs 33\% slower than Eq.~\eqref{eq:surrogate} predicts (range 5--79\%), and $6{\times}6{\times}32$ ($\aspect \approx 5.33$) runs 61\% slower (17--148\%). Conversely, $8{\times}3{\times}48$ runs 36\% faster than predicted (averaged over the eight algorithms with retained runs), and faster than $2{\times}18{\times}32$ at the same $\aspect = 16$. \rev{Orientation matters too: the same dimensions in a different axis order ran up to about 50\% faster or slower at equal rank density (e.g., $6{\times}24{\times}8$ vs.\ $8{\times}6{\times}24$), although $\aspect$ is identical.} Partitions with equal aspect ratio can therefore perform differently, so $\aspect$ alone does not capture all geometric effects. \rev{Pinning down the cause would require} per-link traffic counters, which are not exposed at this scale (Section~\ref{sec:discussion:limitations}). We leave to future work a multivariate model that combines diameter, bisection width, and the mapping of the logical partition onto Tofu's six-dimensional physical coordinates~\cite{ajima2018tofud}, validated against link-level measurements. In its current form, the scalar $\aspect$ already serves as a submission-time screening heuristic that requires no routing knowledge.}

\subsection{Energy Scaling Characteristics}
\label{sec:eval:energy}

\begin{figure}[t]
\centering
\includegraphics[width=\columnwidth]{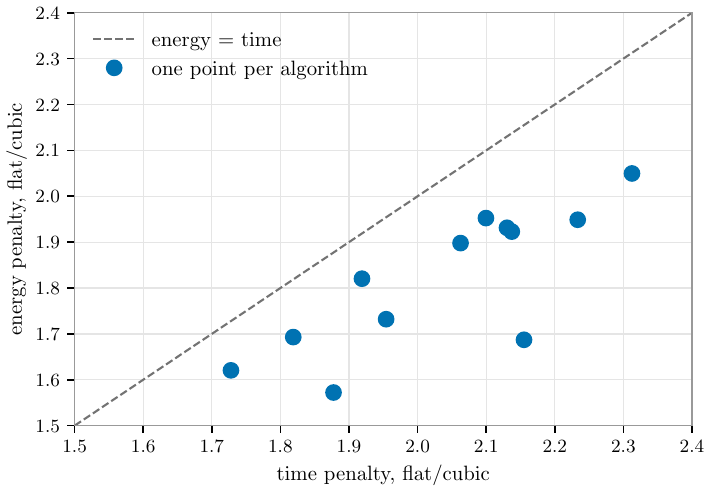}
\caption{Energy penalty versus time penalty of flat relative to
near-cubic partitions, one point per algorithm. Every algorithm falls
below the diagonal: energy grows sub-linearly with time.}
\label{fig:energy}
\end{figure}

Table~\ref{tab:shape} and Figure~\ref{fig:energy} compare flat-to-cubic performance ratios
for time and energy consumption. Energy penalties (\rev{1.57--2.05$\times$, mean 1.82$\times$})
consistently remain below execution time penalties (mean \rev{2.04$\times$}). \rev{On flat partitions, processors spend more time waiting for network communication, which is consistent with a roughly 6\% lower average node power (job energy divided by job elapsed time) than on near-cubic allocations (power ratio 0.92--0.98 across algorithms).} Total
energy calculations integrate all allocated nodes, including idle padding,
identically across both classes. \rev{Energy scales sub-linearly with time because of the lower node power on flat partitions and because fixed per-job phases outside the timed pass add similar energy in both classes.} \rev{Even so,} near-cubic partitions improve both metrics, reducing execution time by roughly half and energy consumption by approximately \rev{45\%} compared to \rev{flat partitions, which default placement assigned most often in our campaign (Table~\ref{tab:default})}.

\subsection{Microbenchmark Corroboration}
\label{sec:eval:micro}

\begin{table}[t]
\centering
\caption{All-to-all microbenchmark: requested versus actually allocated
geometry (recovered from scheduler records) and sustained per-rank
bandwidth at 1\,MiB messages (one rank per node), sorted by allocated
aspect ratio. Default submissions request no shape. \rev{Nodes: allocated; Used: nodes (ranks) that ran the benchmark.}}
\label{tab:micro}
\small
\begin{tabular}{@{}llrrrr@{}}
\toprule
Requested & Allocated & Nodes & \rev{Used} & \aspect & \rev{GiB/s} \\
\midrule
$16{\times}16{\times}8$ & $16{\times}9{\times}16$ & 2,304 & \rev{2,048} & 1.8 & 1.72 \\
(default) & $10{\times}15{\times}8$ & 1,200 & \rev{1,024} & 1.9 & 2.60 \\
$16{\times}8{\times}8$ & $16{\times}9{\times}8$ & 1,152 & \rev{1,024} & 2.0 & 1.97 \\
$32{\times}16{\times}4$ & $4{\times}18{\times}32$ & 2,304 & \rev{2,048} & 8.0 & 0.77 \\
$32{\times}16{\times}2$ & $2{\times}18{\times}32$ & 1,152 & \rev{1,024} & 16.0 & 0.79 \\
$32{\times}16{\times}1$ & $2{\times}18{\times}32$ & 1,152 & \rev{512} & 16.0 & 0.77 \\
$32{\times}32{\times}1$ & $2{\times}33{\times}32$ & 2,112 & \rev{1,024} & 16.5 & 0.86 \\
$32{\times}32{\times}2$ & $2{\times}33{\times}32$ & 2,112 & \rev{2,048} & 16.5 & 0.86 \\
(default) & $2{\times}24{\times}48$ & 2,304 & \rev{2,048} & 24.0 & 0.63 \\
\bottomrule
\end{tabular}
\end{table}

\begin{figure}[t]
\centering
\includegraphics[width=\columnwidth]{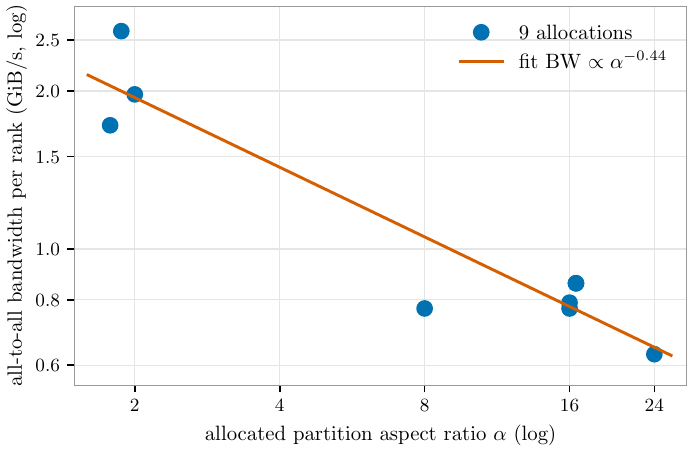}
\caption{All-to-all per-rank bandwidth versus actually allocated aspect ratio for
the nine allocations of Table~\ref{tab:micro}, with the fitted power law
$\mathrm{BW} \propto \aspect^{-0.44}$ ($r = -0.94$). Two allocations
coincide at $\aspect = 16.5$.}
\label{fig:micro}
\end{figure}

\rev{A standalone MPI all-to-all benchmark shows that the interconnect alone reproduces this geometric sensitivity, without any simulator software.} We ran it (1\,MiB messages, 20 timed iterations
after warm-up, one rank per node) across allocations whose \emph{actual}
geometries were recovered from scheduler records; requested and allocated geometries frequently differ \rev{because the scheduler's padding rules override shape requests at unaligned node counts} (\rev{Section~\ref{sec:proposal:mechanism};} Table~\ref{tab:micro}; e.g., a $32{\times}16{\times}4$ request was placed
as $4{\times}18{\times}32$). Across nine allocations \rev{(seven distinct geometries)} \rev{of 1,152 to 2,304 nodes, of which 512 to 2,048 ran the benchmark}, near-cubic geometries
($\aspect \approx 2$) sustained 1.7--2.6\,\rev{GiB/s} per rank, whereas flat allocations
($\aspect = 16$--$24$) sustained 0.63--0.86\,\rev{GiB/s}, a 2.7$\times$ class
difference; the slowest cell (0.63\,\rev{GiB/s}) is an unshaped default request that the
scheduler placed at $2{\times}24{\times}48$. Across all nine allocations, bandwidth falls off as a power law
in aspect ratio with exponent $-0.44$ ($r = -0.94$, Figure~\ref{fig:micro}). \rev{This exponent is close to the application-level $0.39$ in Eq.~\eqref{eq:surrogate}; the small difference is within the uncertainty of a nine-point fit and is in the direction expected if applications dilute network penalties with local computation.}
Although \rev{the number of participating nodes varies by 4$\times$}, making the exponent
indicative rather than exact, this simulator-free measurement reproduces a \rev{similar} systematic power-law dependence on the same hardware platform, \rev{supporting} the network delivery-speed mechanism.

\section{Discussion and Limitations}
\label{sec:discussion}

\subsection{Limitations}
\label{sec:discussion:limitations}

Five experimental limitations qualify these claims. (1)~Using a 1,024-node subset
within a 1,152-node prism creates a slightly irregular communication geometry
compared to the nominal shape; however, the microbenchmark of Section~\ref{sec:eval:micro} \rev{reproduces the same systematic
trend across subsets that use 44--97\% of their allocations, which suggests that this irregularity does not drive the effect}.
(2)~Each configuration cell is timed using a single full-circuit pass
within one job allocation. The warm-up pass agrees with the timed pass to a median difference of \rev{0.6\%, with 25 of the 764 cells differing by more than 5\% (at most 13\%); we did not repeat allocations, so run-to-run variability across allocations is not quantified.}
(3)~\rev{All circuits are fixed-depth representative kernels: deep algorithms use truncated iteration counts, and the Shor-style kernel omits modular exponentiation}; 
\rev{while these kernels exercise the same global state-exchange pattern that production runs repeat (Section~\ref{sec:eval:setup}), they do not reflect full end-to-end algorithmic runtimes lasting hours.} (4)~Link-level hardware counters
are not exposed at this scale, so network-bound behavior is inferred from invariant
traffic volume, geometry-dependent execution times, and the microbenchmark results
of Section~\ref{sec:eval:micro}, rather than \rev{direct link-level measurements}. (5)~The recommended
practice trades queue wait time for execution time: an explicit shape request may wait
longer in job queues than a default submission. 
\rev{In our campaign, compact shape requests showed no such penalty relative to flat ones: among our 1,152-node requests, near-cubic jobs often started ahead of earlier-submitted flat jobs, whereas flat jobs only rarely started ahead of earlier-submitted near-cubic ones. A broader comparison against unshaped submissions and across system load conditions remains future work.}

\subsection{Generalization Beyond Fugaku}
\label{sec:discussion:beyond}

While the numerical constants reported here are specific to Fugaku, the underlying
system-level principles generalize. Any architecture allocating contiguous partitions
on a torus network presents a similar low-dimensional placement variable. Workloads
with placement-invariant communication volumes and \rev{communication-bound} execution should
exhibit systematic geometry penalties governed by system-specific power-law exponents.
The methodology of Sections~\ref{sec:proposal:mechanism}
and~\ref{sec:eval:setup}, including partial node utilization to resolve
alignment constraints, transfers directly to similar environments. On dragonfly or fat-tree interconnects, allocations are \rev{typically non-contiguous and share network links with other jobs, so the placement variable is neither low-dimensional nor isolated from inter-job contention} (Section~\ref{sec:background}); on those systems,
an equivalent surrogate model would score allocations using measured congestion states
rather than static geometry. Nor is the effect specific to quantum
simulation: the microbenchmark of Section~\ref{sec:eval:micro} reproduces
the geometry penalty with a plain MPI all-to-all and no simulator code, so \rev{workloads dominated by global all-to-all or transpose-style collectives, such as parallel three-dimensional FFTs and spectral solvers, are likely to benefit from the same shape request; whether the conditional rank-density effect carries over remains to be tested}. Within quantum simulation itself, because \rev{distributed state-vector simulators generally rely on similar pairwise state-exchange patterns, the geometry penalty should also appear for} GPU-based simulators on torus-like interconnects, even if specific
power-law exponents differ. 
\rev{Furthermore, the emergence of multi-tier accelerated systems, such as NVIDIA GB200 clusters using the Slurm Block Topology plugin~\cite{nvidia2024slurm}, underscores that topology-aware placement remains critical across architectures: when all-to-all exchanges spill outside high-bandwidth NVLink domains into the slower scale-out fabric, workloads suffer sharp bandwidth drops, a placement sensitivity related to the torus geometry penalties observed here.}

\subsection{Implications for Workload Management}
\label{sec:discussion:implications}

These findings indicate that job placement should be treated as a co-design problem
rather than \rev{purely as a utilization problem}. \rev{Current allocation policies appear to treat the padding of large jobs (12.5\% for our 1,024-node submissions) as unmanaged overhead and, consistent with maximizing occupancy, often place such jobs in elongated remnants.} However, for distributed quantum simulation, converting that
same padding into an explicit geometry request improves performance by a factor
of two
\rev{---well above the 20--30\% runtime improvement that Pascual et al.\ found sufficient to compensate for contiguous allocation in fat-tree simulations~\cite{pascual2009job}}. This suggests three operational guidelines for workload managers. First,
submission interfaces should support a \emph{shape intent} specification,
at minimum a flag indicating preference for compact geometry, enabling
schedulers to balance
node utilization against predicted execution slowdowns. Second, schedulers can incorporate
the surrogate model in Eq.~\eqref{eq:surrogate} at queue evaluation time to quantify the
trade-off between immediate execution on flat regions and waiting for compact allocations.
Third, \rev{because energy falls together with execution time on compact partitions (flat partitions consume 1.82$\times$ more energy on average, Section~\ref{sec:eval:energy}), compact placement improves both performance and energy efficiency, so the two objectives need not be traded off in this workload.}
Implementing these policies requires no application modifications and can be integrated
directly into the Quantum--HPC middleware layer discussed in Section~\ref{sec:background}. 
\rev{Regarding the 128 idle nodes incurred under partial use (Section~\ref{sec:proposal:mechanism}), these nodes belong to the job's node-exclusive allocation and cannot be released to other users under the current scheduler. The job itself can still use them: Fujitsu MPI supports simultaneous \texttt{mpiexec} launches on disjoint node sets within one job. Work that generates little network traffic, such as single-node circuit generation or result post-processing, is the natural fit; multi-node tasks are possible but would share Tofu links with the simulation and risk slowing it.}

\section{Conclusion}
\label{sec:conclusion}

\rev{When its node count is set by memory, distributed quantum circuit simulation} is a
network-bound workload whose execution performance is \rev{strongly affected by placement decisions fixed at job submission}. Across \rev{764} controlled configurations
on Fugaku, we demonstrated that these placement decisions follow a \rev{clear} hierarchy: partition geometry acts as a \rev{consistent} factor-of-two performance lever (\rev{1.73--2.31$\times$} across all twelve algorithms under invariant traffic
volume), whereas rank density acts as a secondary, conditional lever \rev{shortening runtime by 11\%} only on compact geometries. \rev{The geometry effect can be modeled to first order} using a scalar predictive metric, the partition aspect ratio, with runtime scaling as $\aspect^{\rev{0.39}}$ \rev{($r = 0.72$)}. The resulting best practice, requesting a near-cubic partition
with 16 ranks per node, improves time-to-solution by roughly 2$\times$ \rev{relative to the flat partitions that default placement mostly assigns, which also consume 1.82$\times$ more energy than near-cubic ones}.

More broadly, these findings indicate that partition geometry
should be managed as a primary scheduling parameter rather than an unmonitored
artifact of utilization packing. Future work should establish formal
confidence intervals through repeated trials, validate the placement
hierarchy and power-law scaling on GPU-based simulators and alternative
torus architectures, and integrate the aspect-ratio surrogate into
production workload managers for cost-aware scheduling. The performance
penalty of ignoring geometry is severe, the predictive model requires
only a single parameter, and the mechanism to \rev{request} controlled
placement can be deployed on unmodified production schedulers today\rev{, provided the allocated shape is checked at job start}.

\section*{Acknowledgment}

First and foremost, we sincerely thank the anonymous reviewers for their in-depth comments. The authors used Anthropic's Claude as a writing and engineering aid when preparing this manuscript. The tool was used to proofread and polish all sections of this manuscript, to write the scripts that generate Figures 2-5 and Tables III-V from the experimental results. All experimental data, findings and conclusions are from the authors and the authors reviewed, verified, and take full responsibility for all content. This work used computational resources of the supercomputer Fugaku provided by RIKEN Center for Computational Science (Project IDs: ra260020 and ra250027) and was partially supported by Japan Science and Technology Agency~(JST) through the Program on Open Innovation Platforms for Industry-academia Co-creation~(COI-NEXT, Grant No.: JPMJPF2221). The MPI version of Qulacs was provided by Fujitsu Research, Fujitsu Ltd.


\balance
\bibliographystyle{IEEEtran}
\bibliography{refs_v5} 

@IEEEtranBSTCTL{IEEEexample:BSTcontrol,
  CTLuse_forced_etal       = "yes",
  CTLmax_names_forced_etal = "6",
  CTLnames_show_etal       = "1"
}

@article{mao2025qgen,
	author =	"Mao, Yikai and Shresthamali, Shaswot and 
			 Kondo, Masaaki",
	title =		"{Q-Gen}: A Parameterized Quantum Circuit Generator",
	journal =	"IEEE Transactions on Quantum Engineering",
	volume =	6,
	pages =	"1--16",
	year =		2025,
	doi =		"10.1109/TQE.2025.3572142"}

@article{jones2019quest,
	author =	"Jones, Tyson and Brown, Anna and 
			 Bush, Ian and Benjamin, Simon C.",
	title =		"{QuEST} and High Performance Simulation of 
			 Quantum Computers",
	journal =	"Scientific Reports",
	volume =	9,
	number =	1,
	pages =		"10736",
	year =		2019,
	doi =		"10.1038/s41598-019-47174-9"}

@misc{pednault2017breaking,
  author       = {Pednault, Edwin and Gunnels, John A. and Nannicini, Giacomo and Horesh, Lior and Magerlein, Thomas and Solomonik, Edgar and Draeger, Erik W. and Holland, Eric T. and Wisnieff, Robert},
  title        = {Pareto-Efficient Quantum Circuit Simulation Using Tensor Contraction Deferral},
  howpublished = {arXiv:1710.05867v4},
  year         = {2020}
}

@article{villalonga2019flexible,
	author =	"Villalonga, Benjamin and Boixo, Sergio and 
			 Nelson, Bron and Henze, Christopher and 
			 Rieffel, Eleanor and Biswas, Rupak and Mandr{\`a}, Salvatore",
	title =		"A Flexible High-Performance Simulator for Verifying 
			 and Benchmarking Quantum Circuits Implemented on 
			 Real Hardware",
	journal =	"npj Quantum Information",
	volume =	5,
	number =	1,
	pages =		"86",
	year =		2019,
	doi =		"10.1038/s41534-019-0196-1"}

@inproceedings{tomesh2022supermarq,
	author =	"Tomesh, Teague and Gokhale, Pranav and Omole, Victory and Ravi, Gokul Subramanian and Smith, Kaitlin N. and Viszlai, Joshua and Wu, Xin-Chuan and Hardavellas, Nikos and Martonosi, Margaret R. and Chong, Frederic T.",
	title =		"{SupermarQ}: A Scalable Quantum Benchmark Suite",
	booktitle =	"Proc. IEEE International Symposium on 
			 High-Performance Computer Architecture (HPCA)",
	pages =		"587--603",
	year =		2022,
	doi =		"10.1109/HPCA53966.2022.00050"}

@article{li2023qasmbench,
	author =	"Li, Ang and Stein, Samuel A. and Krishnamoorthy, Sriram and 
			 Ang, James",
	title =		"{QASMBench}: A Low-Level Quantum Benchmark Suite for 
			 {NISQ} Evaluation and Simulation",
	journal =	"ACM Transactions on Quantum Computing",
	volume =	4,
	number =	2,
	pages =		"1--26",
	year =		2023,
	doi =		"10.1145/3550488"}

@article{chen2018high,
	author =	"Chen, Zhao-Yun and Zhou, Qi and Xue, Cheng and 
			 Yang, Xia and Guo, Guang-Can and Guo, Guo-Ping",
	title =		"64-Qubit Quantum Circuit Simulation",
	journal =	"Science Bulletin",
	volume =	63,
	number =	15,
	pages =		"964--971",
	year =		2018,
	doi =		"10.1016/j.scib.2018.06.007"}

@inproceedings{wu2020full,
	author =	"Wu, Xin-Chuan and Di, Sheng and 
			 Dasgupta, Emma Maitreyee and Cappello, Franck and 
			 Finkel, Hal and Alexeev, Yuri and Chong, Frederic T.",
	title =		"Full-State Quantum Circuit Simulation by Using 
			 Data Compression",
	booktitle =	"Proc. International Conference for High Performance 
			 Computing, Networking, Storage and Analysis (SC)",
	pages =		"1--24",
	year =		2019,
	doi =		"10.1145/3295500.3356155"}

@inproceedings{ajima2018tofud,
	author =	"Ajima, Yuichiro and Kawashima, Takahiro and 
			 Okamoto, Takayuki and Shida, Naoyuki and 
			 Hirai, Kouichi and Shimizu, Toshiyuki and 
			 Hiramoto, Shinya and Ikeda, Yoshiro and 
			 Yoshikawa, Takahide and Uchida, Kenji and 
			 Inoue, Tomohiro",
	title =		"The {Tofu} Interconnect {D}",
	booktitle =	"Proc. IEEE International Conference on Cluster 
			 Computing (CLUSTER)",
	pages =		"646--654",
	year =		2018,
	doi =		"10.1109/CLUSTER.2018.00090"}

@article{suzuki2021qulacs,
  doi = {10.22331/q-2021-10-06-559},
  title = {Qulacs: a fast and versatile quantum circuit simulator for research purpose},
  author = {Suzuki, Yasunari and Kawase, Yoshiaki and Masumura, Yuya and Hiraga, Yuria and Nakadai, Masahiro and Chen, Jiabao and Nakanishi, Ken M. and Mitarai, Kosuke and Imai, Ryosuke and Tamiya, Shiro and Yamamoto, Takahiro and Yan, Tennin and Kawakubo, Toru and Nakagawa, Yuya O. and Ibe, Yohei and Zhang, Youyuan and Yamashita, Hirotsugu and Yoshimura, Hikaru and Hayashi, Akihiro and Fujii, Keisuke},
  journal = {{Quantum}},
  volume = {5},
  pages = {559},
  month = oct,
  year = {2021}
}

@inproceedings{imamura2022mpiqulacs,
  author    = {Tabuchi, Akihiro and Imamura, Satoshi and Yamazaki, Masafumi and Honda, Takumi and Kasagi, Akihiko and Nakao, Hiroshi and Fukumoto, Naoto and Nakashima, Kohta},
  title     = {{mpiQulacs}: A Scalable Distributed Quantum Computer Simulator for {ARM}-based Clusters},
  booktitle = {Proceedings of the 2023 IEEE International Conference on Quantum Computing and Engineering (QCE)},
  pages     = {959--969},
  year      = {2023},
  doi       = {10.1109/QCE57702.2023.00110}
}

@inproceedings{bhatele2013neighborhood,
  author    = {Bhatele, Abhinav and Mohror, Kathryn and Langer, Steven H. and Isaacs, Katherine E.},
  title     = {There Goes the Neighborhood: Performance Degradation due to Nearby Jobs},
  booktitle = {Proceedings of the International Conference on High Performance Computing, Networking, Storage and Analysis (SC '13)},
  pages     = {1--12},
  year      = {2013},
  publisher = {ACM},
  doi       = {10.1145/2503210.2503247}
}

@article{li2017topology,
  author    = {Li, Kangkang and Malawski, Maciej and Nabrzyski, Jarek},
  title     = {Topology-aware Job Allocation in {3D} Torus-based {HPC} Systems with Hard Job Priority Constraints},
  journal   = {Procedia Computer Science},
  volume    = {108},
  pages     = {515--524},
  year      = {2017},
  note      = {{I}nternational Conference on Computational Science (ICCS 2017)},
  doi       = {10.1016/j.procs.2017.05.016}
}

@inproceedings{cuquantum2023,
  author    = {Bayraktar, Harun and Charara, Ali and Clark, David and Cohen, Saul and Costa, Timothy and Fang, Yao-Lung L. and Gao, Yang and Guan, Jack and Gunnels, John and Haidar, Azzam and Hehn, Andreas and Hohnerbach, Markus and Jones, Matthew and Lubowe, Tom and Lyakh, Dmitry and Morino, Shinya and Springer, Paul and Stanwyck, Sam and Terentyev, Igor and Varadhan, Satya and Wong, Jonathan and Yamaguchi, Takuma},
  title     = {{cuQuantum SDK}: A High-Performance Library for Accelerating Quantum Science},
  booktitle = {Proceedings of the 2023 IEEE International Conference on Quantum Computing and Engineering (QCE)},
  pages     = {1050--1061},
  year      = {2023},
  doi       = {10.1109/QCE57702.2023.00119}
}

@article{lazyqr2024,
  author    = {Teranishi, Yusuke and Hiraoka, Shoma and Mizukami, Wataru and Okita, Masao and Ino, Fumihiko},
  title     = {Lazy Qubit Reordering for Accelerating Parallel State-Vector-based Quantum Circuit Simulation},
  journal   = {ACM Transactions on Quantum Computing},
  volume    = {6},
  number    = {4},
  pages     = {27:1--27:33},
  doi       = {10.1145/3748261},
  year      = {2025}
}

@article{multigpunet2025,
  author    = {Brown, W. Michael and Ramesh, Anurag and Lubinski, Thomas and Nguyen, Thien and Bernal Neira, David E.},
  title     = {Multi-{GPU} Quantum Circuit Simulation and the Impact of Network Performance},
  journal   = {Computer Physics Communications},
  volume    = {324},
  pages     = {110126},
  year      = {2026},
  doi       = {10.1016/j.cpc.2026.110126}
}

@inproceedings{pilotquantum2024,
  author    = {Mantha, Pradeep and Kiwit, Florian J. and Saurabh, Nishant and Jha, Shantenu and Luckow, Andre},
  title     = {{Pilot-Quantum}: A Middleware for Quantum-{HPC} Resource, Workload and Task Management},
  booktitle = {Proceedings of the 2025 IEEE 25th International Symposium on Cluster, Cloud and Internet Computing (CCGrid)},
  pages     = {1--10},
  publisher = {IEEE},
  year      = {2025},
  doi       = {10.1109/CCGRID64434.2025.00070}
}

@article{qpusharing2026,
  author    = {Cipollini, Marco and Rizzo, Simone and Iserte, Sergio and Viviani, Paolo and Vitali, Giacomo and Barbieri, Matteo and Bettonte, Gabriella and Boella, Elisabetta and Ganz, Fulvio and Rocco, Roberto and Spina, Orazio and Pe{\~n}a, Antonio J. and Sand{\aa}s, Petter and Colonnelli, Iacopo and Scionti, Alberto and Vercellino, Chiara and Dri, Emanuele and Frassineti, Jonathan and Marzella, Sara and Muratori, Andrea and Ottaviani, Daniele and Terzo, Olivier and Montrucchio, Bartolomeo and Gregori, Daniele},
  title     = {Three Ways to Share a {QPU}: Scheduling Strategies for Hybrid Quantum-{HPC} Applications},
  journal   = {Future Generation Computer Systems},
  volume    = {185},
  pages     = {108699},
  year      = {2026},
  doi       = {10.1016/j.future.2026.108699}
}

@inproceedings{elephantscheduling2025,
  author    = {Viviani, Paolo and Rocco, Roberto and Barbieri, Matteo and Bettonte, Gabriella and Boella, Elisabetta and Cipollini, Marco and Frassineti, Jonathan and Ganz, Fulvio and Marzella, Sara and Ottaviani, Daniele and Rizzo, Simone and Scionti, Alberto and Vercellino, Chiara and Vitali, Giacomo and Terzo, Olivier and Montrucchio, Bartolomeo and Gregori, Daniele},
  title     = {Assessing the Elephant in the Room in Scheduling for Current Hybrid {HPC-QC} Clusters},
  booktitle = {Proceedings of the 55th Annual IEEE/IFIP International Conference on Dependable Systems and Networks Workshops (DSN-W)},
  pages     = {184--187},
  publisher = {IEEE},
  year      = {2025},
  doi       = {10.1109/DSN-W65791.2025.00059}
}

@inproceedings{yang2016bully,
  author    = {Yang, Xu and Jenkins, John and Mubarak, Misbah and Ross, Robert B. and Lan, Zhiling},
  title     = {Watch Out for the Bully! {Job} Interference Study on Dragonfly Network},
  booktitle = {Proceedings of the International Conference for High Performance Computing, Networking, Storage and Analysis (SC '16)},
  pages     = {750--760},
  year      = {2016},
  publisher = {IEEE},
  doi       = {10.1109/SC.2016.63}
}

@inproceedings{jokanovic2015quiet,
  author    = {Jokanovic, Ana and Sancho, Jose Carlos and Rodriguez, German and Lucero, Alejandro and Minkenberg, Cyriel and Labarta, Jesus},
  title     = {Quiet Neighborhoods: Key to Protect Job Performance Predictability},
  booktitle = {IEEE International Parallel and Distributed Processing Symposium (IPDPS)},
  pages     = {449--459},
  year      = {2015},
  doi       = {10.1109/IPDPS.2015.87}
}

@inproceedings{desensi2020slingshot,
  author    = {De Sensi, Daniele and Di Girolamo, Salvatore and McMahon, Kim H. and Roweth, Duncan and Hoefler, Torsten},
  title     = {An In-Depth Analysis of the {Slingshot} Interconnect},
  booktitle = {Proceedings of the International Conference for High Performance Computing, Networking, Storage and Analysis (SC '20)},
  pages     = {1--14},
  year      = {2020},
  publisher = {IEEE},
  doi       = {10.1109/SC41405.2020.00039}
}

@article{deraedt2019massively,
  author    = {De Raedt, Hans and Jin, Fengping and Willsch, Dennis and Willsch, Madita and Yoshioka, Naoki and Ito, Nobuyasu and Yuan, Shengjun and Michielsen, Kristel},
  title     = {Massively Parallel Quantum Computer Simulator, Eleven Years Later},
  journal   = {Computer Physics Communications},
  volume    = {237},
  pages     = {47--61},
  year      = {2019},
  doi       = {10.1016/j.cpc.2018.11.005}
}

@inproceedings{haner2017petabyte,
  author    = {H{\"a}ner, Thomas and Steiger, Damian S.},
  title     = {0.5 Petabyte Simulation of a 45-Qubit Quantum Circuit},
  booktitle = {Proceedings of the International Conference for High Performance Computing, Networking, Storage and Analysis (SC '17)},
  pages     = {1--10},
  year      = {2017},
  publisher = {ACM},
  doi       = {10.1145/3126908.3126947}
}

@article{guerreschi2020intel,
  author    = {Guerreschi, Gian Giacomo and Hogaboam, Justin and Baruffa, Fabio and Sawaya, Nicolas P. D.},
  title     = {{Intel Quantum Simulator}: A Cloud-Ready High-Performance Simulator of Quantum Circuits},
  journal   = {Quantum Science and Technology},
  volume    = {5},
  number    = {3},
  pages     = {034007},
  year      = {2020},
  doi       = {10.1088/2058-9565/ab8505}
}

@inproceedings{cacheblocking2026,
  author    = {Wang, Chuan-Chi and Wang, Yan-Jie and Tu, Chia-Heng and Hung, Shih-Hao},
  title     = {Large-Scale Quantum Circuit Simulation on {HPC} Cluster via Cache Blocking, Boosting, and Gate Fusion Optimization},
  booktitle = {Proceedings of the 55th International Conference on Parallel Processing (ICPP '26)},
  pages     = {89--99},
  year      = {2026},
  publisher = {ACM},
  doi       = {10.1145/3832810.3832819}
}

@misc{qrmi2026,
  author    = {Badts, Thomas and Boyle, Tim and Carvalho, Claudio and C{\'o}rcoles, Antonio and Damin, Andrew and Elisseev, Vadim and Frassineti, Jonathan and Gruber, Daniel and Horii, Hiroshi and Lee, Eun-Kyung and Machin, James and Marzella, Sara and Meller, Mateusz and Milroy, Daniel and Moreau, Matthieu and Ohtani, Munetaka and Ortega-Carrasco, Elisabeth and Oucharek, Doug and Park, Yoonho and Patil, Adarsh and Sahin, Emre M. and Samu, G{\'a}bor and Seelam, Seetharami and Shehata, Amir and Sochat, Vanessa and Thorne, James and Wallis, Oscar and Wennersteen, Aleksander},
  title     = {Examining {QRMI} as a Unified Interface for Quantum-{HPC} Integration},
  howpublished = {arXiv:2607.19591},
  year      = {2026}
}

@inproceedings{pascual2009job,
  author    = {Pascual, Jose Antonio and Navaridas, Javier and Miguel-Alonso, Jose},
  title     = {Effects of Topology-Aware Allocation Policies on Scheduling Performance},
  booktitle = {Job Scheduling Strategies for Parallel Processing (JSSPP 2009)},
  series    = {Lecture Notes in Computer Science},
  volume    = {5798},
  pages     = {138--156},
  publisher = {Springer},
  year      = {2009},
  doi       = {10.1007/978-3-642-04633-9_8}
}

@article{zhou2016improving,
  author    = {Zhou, Zhou and Yang, Xu and Lan, Zhiling and Rich, Paul and Tang, Wei and Morozov, Vitali and Desai, Narayan},
  title     = {Improving Batch Scheduling on {Blue Gene/Q} by Relaxing Network Allocation Constraints},
  journal   = {IEEE Transactions on Parallel and Distributed Systems},
  volume    = {27},
  number    = {11},
  pages     = {3269--3282},
  year      = {2016},
  doi       = {10.1109/TPDS.2016.2528247}
}

@inproceedings{oltchik2020network,
  author    = {Oltchik, Yishai and Schwartz, Oded},
  title     = {Network Partitioning and Avoidable Contention},
  booktitle = {Proceedings of the 32nd ACM Symposium on Parallelism in Algorithms and Architectures (SPAA)},
  pages     = {563--565},
  publisher = {ACM},
  year      = {2020},
  doi       = {10.1145/3350755.3400242}
}

@misc{nvidia2024slurm,
  author       = {Abecassis, Felix and Karakasis, Vasileios and Nabong, Bryan and Wightman, Douglas},
  title        = {Achieving Peak System and Workload Efficiency on {NVIDIA GB200 NVL72} with {Slurm} Block Scheduling},
  howpublished = {NVIDIA Technical Blog},
  month        = may,
  year         = {2026},
  note         = {\url{https://developer.nvidia.com/blog/achieving-peak-system-and-workload-efficiency-on-nvidia-gb200-nvl72-with-slurm-block-scheduling/}}
}

\end{document}